\documentclass{article}
\usepackage{spconf,amsmath,graphicx, url}

\title{A Transformer with Interleaved Self-attention and Convolution for Hybrid Acoustic Models}
\name{Liang Lu}
\address{Microsoft Speech and Language Group\\
\small \tt{liang.lu@microsoft.com}}
\begin{document}
\ninept
\maketitle
\begin{abstract}
Transformer with self-attention has achieved great success in the area of nature language processing. Recently, there have been a few studies on transformer for end-to-end speech recognition, while its application for hybrid acoustic model is still very limited. In this paper, we revisit the transformer-based hybrid acoustic model, and propose a model structure with interleaved self-attention and 1D convolution, which is proven to have faster convergence and higher recognition accuracy. We also study several aspects of the transformer model, including the impact of the positional encoding feature, dropout regularization, as well as training with and without time restriction. We show competitive recognition results on the public Librispeech dataset when compared to the Kaldi baseline at both cross entropy training and sequence training stages. For reproducible research, we release our source code and recipe within the PyKaldi2 toolbox.
\end{abstract}
\section{Introduction}
\label{sec:intro}

Recurrent neural networks (RNNs) with long short-term memory (LSTM)~\cite{hochreiter1997long} units have defined the state-of-the-art large-scale speech recognition since 2014~\cite{sak2014long}. While there have been new types of sequence modeling approaches which are proposed and explored for speech recognition recently,  such as sequence-to-sequence with attention~\cite{Chorowski2015Attention, chan2016listen, lu2015study}, connectionist temporal classification~\cite{graves2006connectionist} and recurrent neural network transducer~\cite{graves2006connectionist}, LSTM-RNNs remains the most popular neural network architectures for learning speech feature representations, although convolutional neural networks (CNNs) with different variants have shown competitive recognition results for some tasks. The key behind the success of RNNs is their capacity to learn temporal correlations in sequential signals through the recurrent connections when the networks are trained with the back-propagation through time (BPTT)~\cite{werbos1990backpropagation} algorithm. However, a well-known weakness of RNNs is the gradient vanishing or explosion problem due to BPTT, and the recurrent connections in RNNs make it challenging to parallelize the computations in both training and inference stages. 

Transformer~\cite{vaswani2017attention}, which relies solely on self-attention to capture the temporal correlations in sequential signals, is a new type of neural network structure for sequence modeling, which has achieved excellent results in machine translation~\cite{vaswani2017attention}, language modeling~\cite{dai2018transformer},  as well as end-to-end speech recognition~\cite{karita2019comparative, dong2018speech}. Self-attention is appealing for sequence modeling in the sense that it can learn long-term correlations by one step of attention operation, while for RNNs, it would take multiple steps in the time space for both forward and backward computation, and noise may accumulate during the process. CNNs, on the other hand, require multiple layers to capture the correlations between the two features which are very distant in the time space, although dilation that uses large strides can reduce the number of layers that is required. While there have been many studies on end-to-end speech recognition using transformers~\cite{karita2019comparative, dong2018speech, sperber2018self, tian2019self, salazar2019self}, their applications for hybrid acoustic models are less well understood. In this paper, we study the more standard transformer for speech recognition within the hybrid framework, and provide further insight to this model through experiments on the Librispeech public dataset.

\section{Related works}

There have been a few studies on transformers for end-to-end speech recognition, particularly for sequence-to-sequence with attention model~\cite{karita2019comparative, dong2018speech, sperber2018self}, as well as transducer~\cite{tian2019self} and CTC models~\cite{salazar2019self}.  In~\cite{karita2019comparative}, the authors compared RNNs with transformers for various speech recognition and synthesis tasks, and obtained competitive or even better results with transformers. However, the key challenge for transformer-based sequence-to-sequence model is to perform online streaming speech recognition, as there is no clear boundary for chunk-wise self-attention. Transformer based transducer~\cite{tian2019self} and CTC model~\cite{salazar2019self} do not have the issue for online speech recognition, however, the results presented in the two studies are not competitive compared the hybrid baseline system from Kaldi~\cite{povey2011kaldi}.  

The work that is closely related to ours is the time restricted self-attention for hybrid acoustic model~\cite{povey2018time}, where the self-attention layer is applied to a chunk of the acoustic frames on top of a time-delay neural network (TDNN) or an LSTM layer. Recently, Han et al.~\cite{han2019multi, han2019state} presented two extensions to this work by using multiple streams of acoustic features to the TDNN layers before the self-attention layer in~\cite{han2019state}, or using multi-stride features to the self-attention layers~\cite{han2019multi}. In fact, the key idea of the two studies is the same, i.e., sample the features using different strides (or sampling rates), and feed the multiple views of the features to the model, assuming that each view contains complementary acoustic information. The only difference is that the multiple views of features are fed into the TDNN layers in~\cite{han2019state} , referred to as multi-stream, while they are fed into the self-attention layers directly in~\cite{han2019multi}, referred to as multi-stride self-attention model. 

In this work, we look at a few other aspects of transformer-based hybrid acoustic models that have not been studied previously. In~\cite{povey2018time, han2019multi, han2019state}, self-attention is only applied in a chunk of acoustic input restricted by a time window, which makes the transformer model easier to train as it does not need to consider very long term correlations. While whole sequence-level self-attention has been applied in sequence-to-sequence models~\cite{karita2019comparative}, hybrid model is different in the sense that it is required to maintain strict frame-level alignments before performing predictions, which may be challenging for a transformer with multiple layers of self-attention as it may reorder the sequence. Furthermore, lower sampling rates are usually used for transformer-based acoustic models, which makes it easier for sequence-level self-attention as the input sequences are much shorter. We propose an interleaved self-attention and convolution structure for transformer model, with the motivation that convolution can learn local feature correlations and maintain the ordering information of the sequence while self-attention can capture long-term correlations.  We show that the model can achieved competitive recognition results when trained with or without time-restriction.  

\section{Transformer}
\label{sec:transformer}

In this section, we review each component in the standard transformer model, and discuss a model structure that is mainly investigated for speech recognition in this work.

\subsection{Self-attention with multiple heads}

The attention mechanism in transformer is technically the same as in the original RNN-based attention model~\cite{bahdanau2014neural}. The key difference is that the query used to compute the attention probability is also from the source sequence, instead of using the decoder hidden state as in the RNN-based attention model~\cite{bahdanau2014neural}. In~\cite{vaswani2017attention}, the authors used the dot-production attention~\cite{luong2015effective} rather than the conventional additive attention~\cite{bahdanau2014neural} in favor of the low computational complexity, which is rewritten here as: 

\begin{align}
\label{eq:att}
\text{Attention}(Q, K, V) = \text{Softmax}\left(\frac{QK^T}{\sqrt{d_k}}\right)V, 
\end{align}
where $Q,K,V$ are referred to the query, key and value according to~\cite{vaswani2017attention}. In transformer, both $Q$ and $K$ are from the source sequence, while in the conventional RNN-based attention model~\cite{bahdanau2014neural}, $Q$ is from the decoder hidden state, and $K$ is from the encoder hidden state. In Eq \eqref{eq:att}, $d_k$ is the dimension of the model, and it is used to scale the dot-product between $Q$ and $K$ in order to smooth the probability distribution returned by the Softmax operation. This is to avoid placing most of the attention probability to a single frame as a result of the dot-product attention, while additive attention does not require such a scaling factor from our experience. 

Another key idea from the transformer paper~\cite{vaswani2017attention} is the multi-head attention mechanism, which performs multiple attention operations in parallel using different model parameters. The output from different attention heads are then concatenated and projected before being fed into the next layer. It can be expressed as
\begin{align}
\text{MultiHead}(Q, K, V) &= [H_1, H_2, \cdots, H_n]W^O \\
\text{where } H_i & = \text{Attention}(QW_i^Q, KW_i^K, VW_i^V) 
\end{align}
where $n$ is the number of attention heads, and $W_i^Q, W_i^K, W_i^V$ are parameters for the $i$-th attention head, and $W^O$ is the projection matrix to reduce the dimension of the concatenated attention vector.

\subsection{Positional encoding}
\label{ssec:pos}

The attention function (\ref{eq:att}) itself does not use the information of the order of the sequence $V$. It is possible that reordering the elements in $V$ can result in the same attention vector after the attention operation, since Eq~\eqref{eq:att} is only a weighted sum of the elements in $V$. To encode the positional information into the model, the authors in~\cite{vaswani2017attention} proposed a sinusoidal function as 
\begin{align*}
PE[t, 2i] &= \sin\left(t /10000^\frac{2i}{d_k}\right) \\
PE[t, 2i+1] &= \cos\left(t/10000^\frac{2i+1}{d_k}\right)
\end{align*}
where $i$ refers to the dimension, and $t$ denotes the time-step. We study the same type of the positional encoding for our speech recognition experiments, by adding $PE[t]$ to the corresponding feature vector at the time step $t$. 
 
\begin{figure}[t]
\small
\centerline{\includegraphics[width=0.25\textwidth]{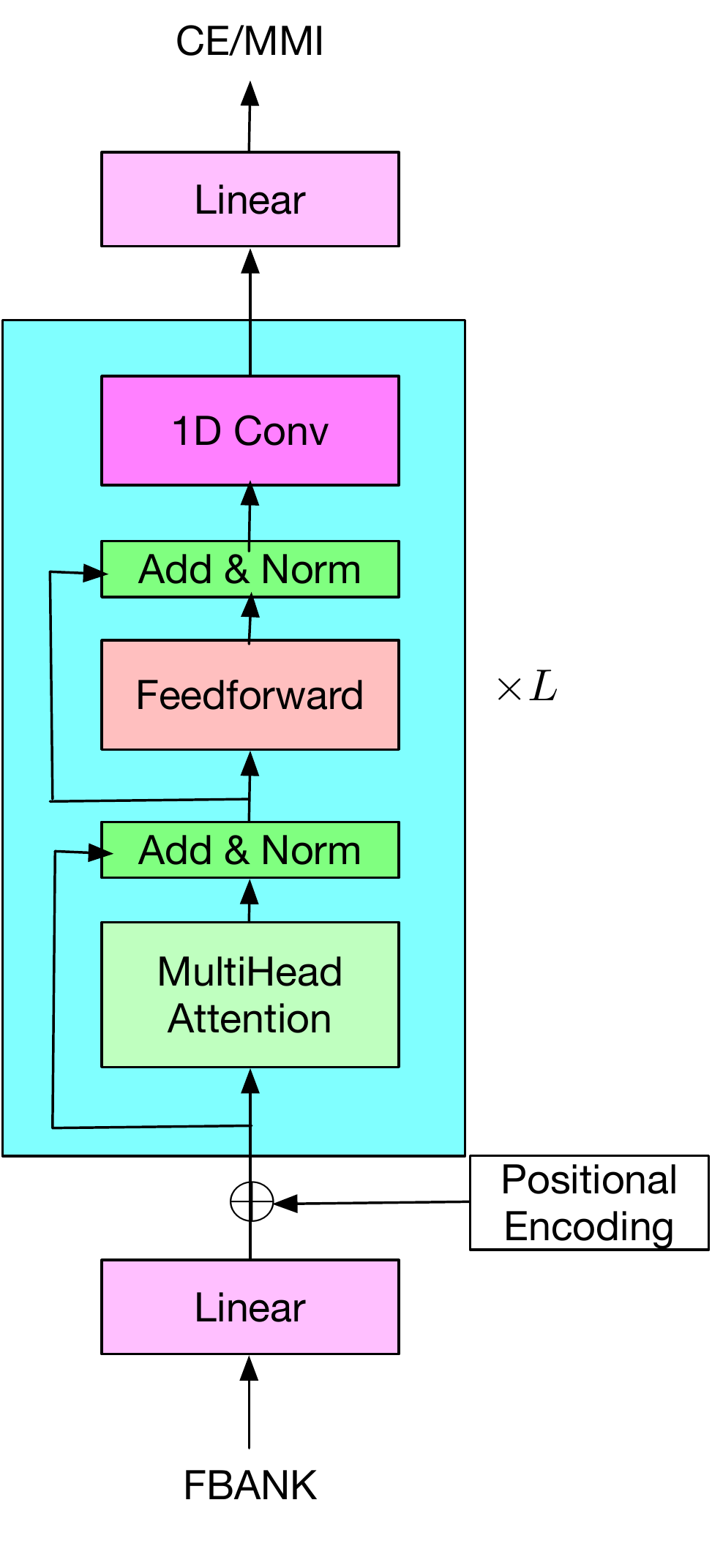}}
\caption{The transformer model structure with interleaved self-attention and 1D convolution. $L$ denotes the number of layers, and {\tt Norm} refers to layer normalization.}  
\label{fig:model}
\vskip -2mm
\end{figure}

\subsection{Interleaved self-attention and convolution}
\label{ssec:conv}

For hybrid models, our preliminary experiments show that transformers with multiple self-attention layers alone is hard to train without time restriction, and it can easily diverge after a few epochs. We hypothesize that this is due to the nature of hybrid models, which are expected to predict the frame-level labels. Hence, they are more sensitive to any reordering or shifting of the acoustic information in the time space compared with sequence-to-sequence models. The positional encoding along may not be able to provide sufficient information to maintain the sequential information in the acoustic sequence (cf. section \ref{ssec:dropout}). In this paper, we propose a transformer model with interleaved 1D convolution and self-attention, with the motivation that the convolution layer can maintain the sequential information of the input sequence, while at the same time, it can learn the local correlations. Self-attention, on the other hand, is expected to capture the global information as the attention is performed as the entire sequence level. The model with interleaved convolution and self-attention has the flexibility to tradeoff the model capacity for learning both local and global information from the input sequence. Same as the standard transformer~\cite{vaswani2017attention}, we also insert the feedforward layer after the multi-head attention. The final model structure is shown in Figure~\ref{fig:model}. It is possible that the feedforward layer is redundant, or its size could be reduced given the 1D convolution layer. We will investigate this aspect in our future work. Table~\ref{tab:size} shows the number of parameters in each component of the transformer model studied in this paper. 

\begin{table}[t]\centering
\caption{The number of parameters in terms of millions (M) for each component in the 6-layer transformer model studied in this paper, where $d_k=512$, the size of feedforward layer is 2048, and the kernel size of the 1D convolution is 3. The feature dimension is 80. }
\label{tab:size}
\footnotesize
\vskip0.15cm
\begin{tabular}{lcc}
\hline 

\hline
Model component         & \#layers &  \#parameters (M)     \\ \hline
Input Linear Layer & 1    & 0.04   \\
MultiHead Attention & 6  & 6.29   \\
Layer Norm & 12 & 0.12 \\
Feedforward & 6  & 12.61   \\
1D-CNN. & 6 & 4.72  \\ 
Output Linear Layer & 1 & 2.96 \\ \hline
Total & & $\sim$26.6 \\ 
\hline

\hline
\end{tabular}
\vskip-3mm
\end{table}

\section{Experiments}
\label{sec:exp}

We performed the experiments using the publicly available Librispeech corpus~\cite{panayotov2015librispeech}, which contains around 960 hours of training data in total. To constrain our research scope, we fixed the depth of the transformer models to be 6 layers, and the dimension of the model $d_k$ in Eq~\eqref{eq:att} to be 512. The number of hidden units in the feedforward layer is 2048. The kernel size for each convolution layer is 3 without stride. The total number of parameters is around 26.6 million. We did some experiments using a smaller model, and the results are worse than what we reported here. We did not train deeper transformer models due to the memory constraint. In our experiments, we used a high frame rate as 100 Hz, i.e., extracting one acoustic frame in every 10 millisecond. This led to long acoustic sequences. As the memory cost of self-attention is in the order of $O(T^2)$, where $T$ is the length of the acoustic sequence, lower frame rate would significantly cut down the memory cost, and enable the training of much deeper transformer models that will studied in our future work. 

In terms of acoustic features, we used 80-dimensional raw log-mel filter-banks (FBANKs), and we did not perform any form of speaker-level feature normalization. Instead, we only applied the utterance-level mean and variance normalization. We used a 4-gram language model for decoding that is released as the part of the corpus, and we used Kaldi~\cite{povey2011kaldi} to build a Gaussian mixture model (GMM) system for bootstrapping. Our transformer acoustic models were trained using the PyKaldi2 toolbox~\cite{lu2019pykaldi2} , which is built on top of Kaldi and PyTorch through the PyKaldi~\cite{can2018pykaldi} wrapper. We used the Adam optimizer~\cite{kingma2014adam} cross entropy (CE) training, and the same learning rate scheduler as in~\cite{vaswani2017attention}. For sequence training, we used the vanilla stochastic gradient decent (SGD) with fixed learning rate.

\subsection{Results of positional encoding and dropout}
\label{ssec:dropout}

We first evaluated the positional encoding discussed in section~\ref{ssec:pos} and dropout training for the transformer model. Results are given in Table~\ref{tab:pos}. Unlike the observations in the area of machine translation, positional encoding did not make a big difference in terms of recognition accuracies for our transformer models. One possible reason is that we have used 1D convolution, which has encoded some sequential information in to the model. Another possible reason is that the dynamic range of the positional encoding is much smaller compared to the output of the first linear layer in Figure~\ref{fig:model}. During the model training, the information in the positional encoding may be ignored. We performed a sanity check by removing the positional encoding when evaluating a transformer model trained {\it with} positional encoding, and obtained results which are only around 0.1\% worse absolute. In the future, we shall investigate if it would make a difference after scaling the positional encoding features to the same dynamic range as the acoustic feature after the linear projection layer. 

As for dropout training, it was pointed out in~\cite{vaswani2017attention} that transformer model for sequence to-sequence ASR may suffer from overfitting easily, and regularization such as dropout is important to address such kind of issue. In our experiments, we also observed the overfitting problem, and our models were usually trained for about 8 - 10 epochs. However, dropout is not effective for our model, and it only slightly improved the recognition accuracy. It may be due to the convolution layers used in our model, as according to our experience, dropout does not work well for CNNs. While other regularization approach may be applicable including adding data noise, in our future work, we are more interested in evaluating the model with very large amount of our internal training data to see if the transformer model is still sensitive to the overfitting problem. 

\begin{table}[t]\centering
\caption{Results of the transformer model with or without positional encoding denoted as {\tt POS}, and dropout regularization denoted as {\tt DP}.}
\label{tab:pos}
\footnotesize
\vskip0.15cm
\begin{tabular}{l|cccccc}
\hline 

\hline
 & & &  \multicolumn{2}{c}{dev} & \multicolumn{2}{c}{test} \\
Model          & POS & DP &  clean       &  other & clean & other \\ \hline
  & $\times$  & 0.1 & 4.5 & 11.5 &  4.9 & 11.8 \\
Transformer & $\surd$  & 0.1 &  4.4 & 11.7 &  4.9 & 11.9 \\
 & $\times $ & 0    & 4.5 & 11.8 & 5.0 & 12.1  \\
 & $\times$ & 0.2  & 4.6 & 11.9 &  5.0  & 12.2 \\

\hline

\hline
\end{tabular}
\vskip-3mm
\end{table}

\begin{table}[t]\centering
\caption{Results of the transformer model with or without 1D convolution. {\tt Norm} refers to the layer normalization before output linear layer.}
\label{tab:conv}
\footnotesize
\vskip0.15cm
\begin{tabular}{l|cccccc}
\hline 

\hline
 & &  &  \multicolumn{2}{c}{dev} & \multicolumn{2}{c}{test} \\
Model          & Conv & {\tt Norm} &  clean       &  other & clean & other \\ \hline
  & $\times$  & $\surd$ & 5.1 & 13.0 &  5.8  & 13.4 \\
Transformer & $\surd$  & $\surd$ &  4.4 & 11.6 &  4.9 & 12.0 \\
 & $\surd $ & $\times$    & 4.5 & 11.5 & 4.9 & 11.8  \\

\hline

\hline
\end{tabular}
\vskip-3mm
\end{table}

\subsection{1D convolution and layer normalization}
\label{ssec:conv}

We then evaluated the impact of the 1D convolution layers in our transformer model by removing all the convolution layers. This corresponds to a vanilla transformer as in~\cite{vaswani2017attention}. We still added the positional encoding feature to the inputs since the sequential information from the convolution layers is no longer available. We also have to insert another layer normalization to the output of the transformer before the output linear layer to stabilize the training. Otherwise, the training diverges quickly after one or two epochs in our experiments. The results are given in Table~\ref{tab:conv}, which shows that the recognition errors are much higher when the model does not have the convolution layers. Besides, without the convolution layers, the convergence of the model during training also became much slower, which demonstrates that the convolution layers are helpful for both convergence in model training and the recognition accuracy. For a fair comparison, we also added another layer normalization to the model with convolutions. While it can further speed up the convergence, it does not improve the recognition accuracy further. In the future, we shall compare the interleaved combination of self-attention and convolution to the sequential combination of self-attention and TDNN~\cite{povey2018time, han2019multi, han2019state}, which is a special type of 1D convolution with subsampling.  

\begin{table}[t]\centering
\caption{Results of the transformer model trained with or without the time restriction.}
\label{tab:time}
\footnotesize
\vskip0.15cm
\begin{tabular}{l|ccccc}
\hline 

\hline
& & \multicolumn{2}{c}{dev} & \multicolumn{2}{c}{test} \\
Model          & window &  clean       &  other & clean & other \\ \hline
 & $[-\infty, \infty]$ & 4.4 & 11.6 &  4.9 & 12.0 \\
Transformer & $[-\infty, 0]$ & 4.6 & 11.8 &  5.1 & 12.5  \\
 & $[-\infty, 12]$ & 4.5 & 12.1 &  5.0  & 12.4  \\
 & $[-\infty, 24]$ & 4.7 & 12.1 &  5.0  & 12.7  \\

\hline

\hline
\end{tabular}
\vskip-3mm
\end{table}

\subsection{With or without time restriction}

In the previous experiments, we performed self-attention across the whole input sequence. This corresponds to the offline model as the whole sequence need to be visible before the attention operation. For online streaming speech recognition, we can simply apply to a time restriction window to the self-attention layer, which is the same as the study in~\cite{povey2018time}. However, this is very challenging for sequence-to-sequence model based on transformer as the boundary for each output token is unclear. For hybrid models, the latency is controllable by adjusting the size of the time restriction window. In our implementation, we still take the whole sequence as the input, but mask out the frames which are outside the time restriction window, i.e., the attention probabilities of those frames are set to be zero during training.  

Follow the convention, we denote $[-l, r]$ as the attention time window, where $l$ and $r$ are the sizes of left and right context. When $l$ is $\infty$, it means that we perform self-attention up to the start of the acoustic sequence, while $r=\infty$ means the attention operation spans to the end of the sequence. The offline model corresponds to the time window of $[-\infty, \infty]$ as in Table~\ref{tab:time}. Note that, each 1D convolution layer looks ahead 1 frame since the kernel size is 3 without stride (e.g., $[-1, 1]$). The time window in Table~\ref{tab:time} is only for the attention operation, and the corresponding latency should have another 6 frames overhead from the convolution layers. In addition, the numbers in Table~\ref{tab:time} refer to the accumulated attention window. If the time window for each self-attention layer is $[-\infty, 2]$, then the total accumulated time window for 6 self-attention layers would be $[-\infty, 12]$. 

For faster convergence, we used the transformer models with one more layer normalization as in section \ref{ssec:conv}, although the offline results on the {\tt dev-other} and {\tt test-other} evaluation sets are slightly worse. The table shows that when we limited the future context information for the transformer model, we obtained slightly worse results. However, contrary to our expectations, when we increased the right context size, we did not achieve higher recognition accuracy, although the CE losses were significantly reduced, e.g., from $\sim$0.78 for the model with the attention window of $[-\infty, 0]$ to $\sim$0.70 for the model with the attention window of $[-\infty, 24]$ from our setups. Although the convolution layers have already looked ahead 6 frame in total, we believe the future context information should still be helpful. We hypothesis that this may be due to the multi-head attention. If one or more attention heads are placed at around the end of the time window, i.e., focusing on the future context information more than it should, the information from those time steps can help to reduce the training loss, but it may not be able to generalize well. To understand deeper about this results, we will replicate the set of experiments with different numbers of attention heads, and also perform experiments on some other datasets in our future work. 

\subsection{Sequence training results} 
\label{ssec:comp}

In Table~\ref{tab:comp}, we show the sequence training results of the transformer model trained with the maximum mutual information (MMI) criterion. We followed the traditional lattice-based sequence training approach, and the lattices were generated on-the-fly as implemented in PyKaldi2. We used a CE trained model as the seed model, and then trained the model with MMI using the vanilla SGD optimizer. We fixed learning rate as $5\times10^{-5}$, and to avoid overfitting, we applied the CE regularization with weight as $0.2$. The model was converged in less than 1 epoch. Table~\ref{tab:comp} shows that we obtained larger improvements on the noisy test sets ({\tt dev-other, test-other}). Our results are comparable to the results of the TDNN system in Kaldi\footnote{\url{https://github.com/kaldi-asr/kaldi/blob/master/egs/librispeech/s5/RESULTS}}, which is a well-tuned hybrid system. In fact, the TDNN system applied the speed perturbation~\cite{ko2015audio} for data argumentation and i-vector based speaker adaptive training, while in our system, we only used the raw log-mel filterbank features without using any speaker-level information. From that sense, our results are very competitive. Han et al.~\cite{han2019multi, han2019state} achieved better results by using multi-stream and multi-stride features on top of the TDNN system, which are also applicable to our system, and will be investigated in the future. 

\section{Conclusion}
\label{sec:conc}

While transformer has been very successful in the area of nature language processing, its application to speech recognition is mostly within the end-to-end architecture. We are more interested in transformers for hybrid acoustic models as there is no theoretical issues for online streaming speech recognition. In this paper, we have presented a transformer model with interleaved self-attention and convolution for hybrid acoustic modeling, although this  structure may be also applicable to end-to-end models. We have showed that the convolutional layers can improve the recognition accuracy with faster convergence compared to the model with self-attention layers only. We have also investigated several other aspects of the model including the impact of the positional encoding feature, dropout regularization as well training with and without the time restriction. Our work is an addition to the current study of self-attention for hybrid models with a sequential TDNN and self-attention architecture trained with time restriction only. For our future works, we shall study training much deeper transformer with low frame rate to get rid of the GPU memory constraint, as well as evaluate the model in the setting with a very large amount of training data. 

\begin{table}[t]\centering
\caption{Sequence training results and comparison to a baseline hybrid system.}
\label{tab:comp}
\footnotesize
\vskip0.15cm
\begin{tabular}{l|ccccc}
\hline 

\hline
& & \multicolumn{2}{c}{dev} & \multicolumn{2}{c}{test} \\
Model          & Criterion &  clean       &  other & clean & other \\ \hline
           & CE  & 4.4 & 11.6 &  5.0 & 12.1 \\
TDNN & sMBR  & 4.1 & 11.1 &  4.6 & 11.3 \\
         & LFMMI  & 3.9 & 10.4 &  4.3 & 10.8 \\ \hline
Transformer & CE& 4.5 & 11.5 &  4.9 & 11.8 \\
 & MMI &4.3  & 10.7 & 4.6 &11.1  \\

\hline

\hline
\end{tabular}
\vskip-3mm
\end{table}

\bibliographystyle{IEEEbib}
\bibliography{bibtex}

\begin{thebibliography}{10}

\bibitem{hochreiter1997long}
Sepp Hochreiter and J{\"u}rgen Schmidhuber,
\newblock ``Long short-term memory,''
\newblock {\em Neural computation}, vol. 9, no. 8, pp. 1735--1780, 1997.

\bibitem{sak2014long}
Hasim Sak, Andrew~W Senior, and Fran{\c{c}}oise Beaufays,
\newblock ``Long short-term memory recurrent neural network architectures for
  large scale acoustic modeling.,''
\newblock in {\em Proc. INTERSPEECH}, 2014.

\bibitem{Chorowski2015Attention}
Jan~K Chorowski, Dzmitry Bahdanau, Dmitriy Serdyuk, Kyunghyun Cho, and Yoshua
  Bengio,
\newblock ``Attention-based models for speech recognition,''
\newblock in {\em Advances in Neural Information Processing Systems}, 2015, pp.
  577--585.

\bibitem{chan2016listen}
William Chan, Navdeep Jaitly, Quoc Le, and Oriol Vinyals,
\newblock ``Listen, attend and spell: A neural network for large vocabulary
  conversational speech recognition,''
\newblock in {\em Proc. ICASSP}. IEEE, 2016, pp. 4960--4964.

\bibitem{lu2015study}
Liang Lu, Xingxing Zhang, Kyunghyun Cho, and Steve Renals,
\newblock ``A study of the recurrent neural network encoder-decoder for large
  vocabulary speech recognition,''
\newblock in {\em Proc. Interspeech}, 2015.

\bibitem{graves2006connectionist}
Alex Graves, Santiago Fern{\'a}ndez, Faustino Gomez, and J{\"u}rgen
  Schmidhuber,
\newblock ``Connectionist temporal classification: labelling unsegmented
  sequence data with recurrent neural networks,''
\newblock in {\em Proc. ICML}. ACM, 2006, pp. 369--376.

\bibitem{werbos1990backpropagation}
Paul~J Werbos et~al.,
\newblock ``Backpropagation through time: what it does and how to do it,''
\newblock in {\em Proceedings of the IEEE}, 1990, vol.~78, pp. 1550--1560.

\bibitem{vaswani2017attention}
Ashish Vaswani, Noam Shazeer, Niki Parmar, Jakob Uszkoreit, Llion Jones,
  Aidan~N Gomez, {\L}ukasz Kaiser, and Illia Polosukhin,
\newblock ``Attention is all you need,''
\newblock in {\em Advances in Neural Information Processing Systems}, 2017, pp.
  5998--6008.

\bibitem{dai2018transformer}
Zihang Dai, Zhilin Yang, Yiming Yang, William~W Cohen, Jaime Carbonell, Quoc~V
  Le, and Ruslan Salakhutdinov,
\newblock ``Transformer-xl: Language modeling with longer-term dependency,''
\newblock in {\em Proc. ICLR}, 2019.

\bibitem{karita2019comparative}
Shigeki Karita, Nanxin Chen, Tomoki Hayashi, Takaaki Hori, Hirofumi Inaguma,
  Ziyan Jiang, Masao Someki, Nelson Enrique~Yalta Soplin, Ryuichi Yamamoto,
  Xiaofei Wang, et~al.,
\newblock ``A comparative study on transformer vs rnn in speech applications,''
\newblock in {\em arXiv preprint arXiv:1909.06317}, 2019.

\bibitem{dong2018speech}
Linhao Dong, Shuang Xu, and Bo~Xu,
\newblock ``Speech-transformer: a no-recurrence sequence-to-sequence model for
  speech recognition,''
\newblock in {\em Proc. ICASSP}. IEEE, 2018, pp. 5884--5888.

\bibitem{sperber2018self}
Matthias Sperber, Jan Niehues, Graham Neubig, Sebastian St{\"u}ker, and Alex
  Waibel,
\newblock ``Self-attentional acoustic models,''
\newblock in {\em arXiv preprint arXiv:1803.09519}, 018.

\bibitem{tian2019self}
Zhengkun Tian, Jiangyan Yi, Jianhua Tao, Ye~Bai, and Zhengqi Wen,
\newblock ``Self-attention transducers for end-to-end speech recognition,''
\newblock in {\em arXiv preprint arXiv:1909.13037}, 2019.

\bibitem{salazar2019self}
Julian Salazar, Katrin Kirchhoff, and Zhiheng Huang,
\newblock ``Self-attention networks for connectionist temporal classification
  in speech recognition,''
\newblock in {\em Proc. ICASSP}. IEEE, 2019, pp. 7115--7119.

\bibitem{povey2011kaldi}
D~Povey, A~Ghoshal, G~Boulianne, L~Burget, O~Glembek, N~Goel, M~Hannemann,
  P~Motl{\i}cek, Y~Qian, P~Schwarz, J~Silovsk\'y, G~Semmer, and K~Vesel\'y,
\newblock ``{The Kaldi speech recognition toolkit},''
\newblock in {\em Proc. ASRU}, 2011.

\bibitem{povey2018time}
Daniel Povey, Hossein Hadian, Pegah Ghahremani, Ke~Li, and Sanjeev Khudanpur,
\newblock ``A time-restricted self-attention layer for asr,''
\newblock in {\em Proc. ICASSP}. IEEE, 2018, pp. 5874--5878.

\bibitem{han2019multi}
Kyu~J Han, Jing Huang, Yun Tang, Xiaodong He, and Bowen Zhou,
\newblock ``Multi-stride self-attention for speech recognition,''
\newblock in {\em Proc. INTERSPEECH}, 2019, pp. 2788--2792.

\bibitem{han2019state}
Kyu~J Han, Ramon Prieto, Kaixing Wu, and Tao Ma,
\newblock ``State-of-the-art speech recognition using multi-stream
  self-attention with dilated 1d convolutions,''
\newblock in {\em arXiv preprint arXiv:1910.00716}, 2019.

\bibitem{bahdanau2014neural}
Dzmitry Bahdanau, Kyunghyun Cho, and Yoshua Bengio,
\newblock ``Neural machine translation by jointly learning to align and
  translate,''
\newblock in {\em Proc. ICLR}, 2015.

\bibitem{luong2015effective}
Minh-Thang Luong, Hieu Pham, and Christopher~D Manning,
\newblock ``Effective approaches to attention-based neural machine
  translation,''
\newblock in {\em arXiv preprint arXiv:1508.04025}, 2015.

\bibitem{panayotov2015librispeech}
Vassil Panayotov, Guoguo Chen, Daniel Povey, and Sanjeev Khudanpur,
\newblock ``Librispeech: an asr corpus based on public domain audio books,''
\newblock in {\em Proc. ICASSP}. IEEE, 2015, pp. 5206--5210.

\bibitem{lu2019pykaldi2}
Liang Lu, Xiong Xiao, Zhuo Chen, and Yifan Gong,
\newblock ``Pykaldi2: Yet another speech toolkit based on kaldi and pytorch,''
\newblock in {\em arXiv preprint arXiv:1907.05955}, 2019.

\bibitem{can2018pykaldi}
Dogan Can, Victor~R Martinez, Pavlos Papadopoulos, and Shrikanth~S Narayanan,
\newblock ``Pykaldi: A python wrapper for kaldi,''
\newblock in {\em Proc. ICASSP}. IEEE, 2018, pp. 5889--5893.

\bibitem{kingma2014adam}
Diederik~P Kingma and Jimmy Ba,
\newblock ``Adam: A method for stochastic optimization,''
\newblock in {\em arXiv preprint arXiv:1412.6980}, 2014.

\bibitem{ko2015audio}
Tom Ko, Vijayaditya Peddinti, Daniel Povey, and Sanjeev Khudanpur,
\newblock ``Audio augmentation for speech recognition,''
\newblock in {\em Proc. INTERSPEECH}, 2015.

\end{thebibliography}

\end{document}